\documentclass[aps,prl,reprint,groupedaddress]{revtex4-1}

\usepackage{graphicx,amsmath,color,amsfonts, amsmath, bm}

\usepackage[colorlinks=true]{hyperref}  
\hypersetup{
    bookmarks=true,         
    unicode=false,          
    pdftoolbar=true,        
    pdfmenubar=true,        
    pdffitwindow=false,     
    pdfstartview={FitH},    
    pdftitle={},    
    pdfauthor={},     
    pdfsubject={},   
    pdfcreator={},   
    pdfproducer={}, 
    pdfkeywords={} {} {}, 
    pdfnewwindow=true,      
    colorlinks=true,       
    linkcolor=blue, 
    citecolor=blue,        
    filecolor=magenta,      
    urlcolor=blue,           
        breaklinks=true
}

\newcommand{\be}{\begin{equation}}
\newcommand{\ee}{\end{equation}}
\newcommand{\bea}{\begin{eqnarray}}
\newcommand{\eea}{\end{eqnarray}}

\begin{document}

\title{Turbulent relaxation after a quench in the Heisenberg model}

\author{Joaquin F. Rodriguez-Nieva}
\affiliation{Department of Physics, Stanford University, Stanford, CA 94305, USA}


\begin{abstract}

  We predict the emergence of turbulent scaling in the quench dynamics of the two-dimensional Heisenberg model for a wide range of initial conditions and model parameters. In the isotropic Heisenberg model, we find that the spin-spin correlation function exhibits universal scaling consistent with a turbulent  energy cascade. When the spin rotational symmetry is broken by an easy-plane exchange anisotropy, we find a dual cascade of energy and an emergent conserved charge associated to transverse magnetization fluctuations. The scaling exponents are estimated analytically and agree with numerical simulations using phase space methods. We also define the space of initial conditions (as a function of energy, magnetization, and spin number $S$)  that lead to a turbulent cascade. The universal character of the cascade, insensitive to microscopic details or initial conditions, suggests that turbulence in spin systems can be broadly realized in cold atom and solid-state experiments. 
  
\end{abstract}



\maketitle

\begin{figure}[b]
\includegraphics[scale = 1.0]{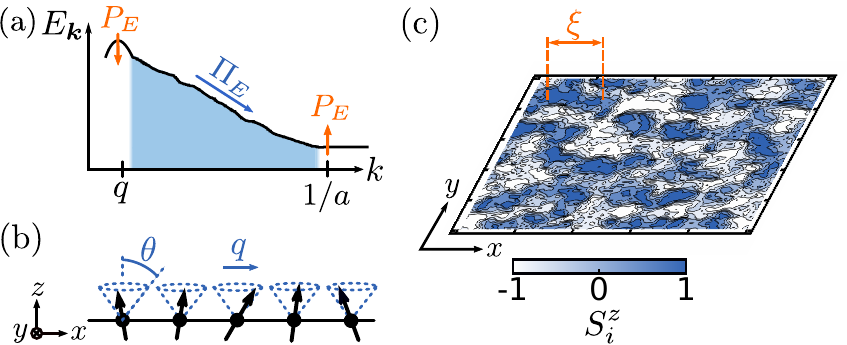}
  \caption{(a) Turbulent states are characterized by fluxes of conserved charges ({\it e.g.}, the energy flux $\Pi_E$ induced by pumping energy at a rate $P_E$) ranging from the wavevector $q$ of the drive to a dissipative lengthscale $a$. In this range, $\Pi_E$ governs the scaling of the energy distribution $E_{\bm k}$. (b) We consider a spin spiral parametrized by a wavevector $q$ and angular amplitude $\theta$ as the initial state. (c) Real space snapshop of $ S_i^z(t) $ at the onset of the turbulent relaxation, $t/\tau_{*} = 10$, shown for a single semiclassical realization with energy $E/N=J/4$ and zero net magnetization[$1/\tau_* = JS(qa\sin\theta)^2$]. Indicated with a bar is the correlation length $\xi\approx 1/q$.} 
  \label{fig:schematics}
\end{figure}

Systems far from thermodynamic equilibrium can exhibit universal dynamics {\it en route} to thermalization. Examples of such phenomena arise when a system is quenched close to a critical point ({\it i.e.}, ageing\cite{2005ageingcalabrese}) or deep in a broken symmetry phase ({\it i.e.}, coarsening\cite{1994Bray}). Turbulence is a different instance of scaling out of equilibrium that can emerge even in the absence of a critical point or long range order\cite{bookzakharov,booknazarenko}. In its simplest form, an external drive at short wavevectors gives rise to a steady-state flux of conserved charges that span many lengthscales up to some dissipative UV scale, see Fig.\ref{fig:schematics}(a). Within this broad lengthscale range---the inertial range---, these fluxes govern the scaling of experimentally relevant correlations. As a result, turbulent states are specified by {\it fluxes} of conserved charges, in contrast to thermal states which are specified by thermodynamic potentials. Importantly, the same scaling can also emerge in an intermediate-time but long-lived prethermal regime after quenching an isolated system, reflecting that the turbulent scaling is intrinsic to the system rather than a feature of the drive. 

Turbulence in quantum systems has been broadly discussed in the context of Bose-Einstein condensates (BECs), both in theory\cite{2002vinenreview,2005prlgardiner,2009tsubotareview,2006pragardiner,2005prlgardiner,2007tsubota,2010prlvelocitystatistics,2010prltsubota,2013gasenzer,2019gasenzer} and experiments\cite{2001prlketterle,2006naturestamperkurn,2008prlstamperkurn,2009prlamoturbulence,2013becturbulenceexp,2016turbulencehadzibabic,2017praspinturbulence,2019turbulencehadzibabic}. The typical scenario is to drive a BEC across a dynamic instability which generates a complex network of vortices, the topological defects of the broken U(1) phase\cite{2005prlstirredbec,2012pratsubota2,2014stirringbec}. Because vortices are energetically stable and long-lived, they play a central role in BEC turbulence\cite{footnote_novortices,2015bogoliubovturbulence} and give rise to rich physics, from Kolmogorov scaling resembling hydrodynamic turbulence\cite{2002vinenreview} to Kelvin wave cascades\cite{2009turbulencelargescale,2011kelvinwaves} to self-similar relaxation\cite{2008berges,2015bergesreview,2015asier,2018natureuniversality1,2018natureuniversality2,2018natureuniversality3,2020hadzibabicuniversality} to connections with holography\cite{2013becholography}.

Here we inquire about the feasibility of realizing turbulent relaxation---and characterize its universal properties---in isolated spin systems, with a special focus on the Heisenberg model in dimension $d=2$. The $d=2$ case is specially interesting as the absence of a phase transition and long range order precludes scaling due to ageing or coarsening. We consider generic initial conditions with arbitrary magnetization and wavevectors, analytically derive the scaling exponents in the limit of large spin number $S$, and numerically show the robustness of the scaling for finite $S$. Our results reveal that spin systems can host a turbulent regime that is qualitatively distinct from BECs in several important ways. Crucially, the key ingredient in BEC turbulence---vortices---is absent. In addition, our results also differ from turbulence in the spin sector of spinor BECs\cite{2012pratsubota1,2016spinturbulence}: spinor BECs host conservation laws that have no analogue in spin models, {\it i.e.}, particle number and momentum. As a result, spin turbulence in spinor BECs is coupled to orbital turbulence and, typically, orbital turbulence dominates the dynamics\cite{2012pratsubota1} except for specific (low energy) initial conditions\cite{2016spinturbulence}. Finally, quantum fluctuations can play a more prominent role in spin systems given the typically small local Hilbert space, therefore restricting the phase space of initial states that lead to a turbulent cascade, as we discuss next. 

\noindent{\bf Phenomenology \& regimes.}---Here we discuss a simple phenomenological picture for two-dimensional spin turbulence which will be used below to derive scaling exponents. Let us consider the relaxation of an excited state with a characteristic wavevector $q$ and average local magnetization $S\sin\theta$, see Fig.\ref{fig:schematics}(b), evolved under the isotropic Heisenberg model with local exchange coupling $J$, spin number $S$, and lattice constant $a$. After an initial transient time on the order of $\tau_* = [JS(qa\sin\theta)^2]^{-1}$, dephasing leads to a non-equilibrium distribution of spin fluctuations with correlation length $\xi \approx 1/q$ quantifying the characteristic size of spin textures\cite{1962skyrme}, see Fig.\ref{fig:schematics}(c). 
The excess energy concentrated at wavevectors $q$ will trigger a turbulent cascade in which energy will be transported incoherently to the UV by quasiparticles with wavevectors $|\bm k|\gtrsim q$. 

Turbulence requires dissipationless transport of charges across many lengthscales (which implies that the Reynold's number is large). Of special importance in cold atomic gases and solid-state materials are quantum dissipative processes which can be prominent at small $S$. One {\it ad hoc} measure of such processes is the ratio between classical fluctuations and quantum fluctuations which, as we explain below, is given by 
\be
   {\cal Q} = \frac{S\sin^2\theta}{(qa)^d}. 
   \label{eq:reynolds}
\ee
When ${\cal Q} \lesssim 1$, we find overdamped dynamics and absence of turbulence. When ${\cal Q} \gg 1$, we find universal scaling in the spin-spin correlation function consistent with wave turbulence, {\it i.e.}, weakly-coupled incoherent waves. The global magnetization of the initial state (parametrized by $\theta$) determines whether correlations are isotropic and all components of magnetization exhibit scaling or, instead, whether only the transverse spin components exhibit scaling.

\noindent{\bf Microscopic model and conserved fluxes.}---We consider the two-dimensional Heisenberg model on a square lattice with short range interactions: 
\be
\hat{H} = -J\sum_{\langle ij\rangle}\hat{S}_i^x \hat{S}_j^x+\hat{S}_i^y \hat{S}_j^y + \Delta\hat{S}_i^z\hat{S}_j^z.
\label{eq:Hxxz}
\ee
Each lattice site has a spin $S$ degrees of freedom and we assume periodic boundary conditions. We begin analyzing the isotropic point ($\Delta=1$) and, after this case has been discussed, extension of the results to the anisotropic case will be straightforward. Starting from a spin spiral, 
\be
\langle \hat{S}_i^\pm\rangle = S\sin\theta e^{\pm i {\bm q}\cdot{\bm r}_i}, \quad \langle \hat{S}_i^z\rangle =S \cos\theta,
\label{eq:initial}
\ee
we study the evolution of magnetization fluctuations through the equal time spin-spin correlation function ${\cal C}_{\bm k}^\alpha(t) = \langle \hat{S}_{-\bm k}^\alpha(t) \hat{S}_{\bm k}^\alpha(t)\rangle$ for $\alpha=x,y,z$. The parameters $\bm q$ and $\theta$ control the energy and total magnetization of the initial state. Triggering turbulence from a spiral state resembles the typical scenario in BEC turbulence\cite{2015PRX-babadi}: in both cases, a dynamic instability gives rise to exponential growth of classical fluctuations on timescales much shorter than the thermalization time and, subsequently, turbulent scaling emerges\cite{2020spinspiral}. 

Because the energy operator overlaps with $\hat{S}_{-\bm k}^\alpha\hat{S}_{\bm k}^\alpha$ for all $\bm k$, the time evolution of ${\cal C}_{\bm k}^\alpha(t)$ is constrained by the flow of energy in $\bm k$-space. In addition, the system may exhibit emergent conserved quantities---quantities which are conserved statistically rather than microscopically---that may also constrain the evolution of ${\cal C}_{\bm k}^\alpha(t)$ and give rise to multiple scaling exponents that characterize different regions of phase space. Notably, the amplitude of transverse magnetization $\hat{N} = \sum_i (\hat{S}_i^x)^2+(\hat{S}_i^y)^2$ may become statistically conserved in several scenarios. One example is when the system is close to the ferromagnetic ground state. In this case, the probability of having two spin flips on the same site is negligible and the picture of a weakly-interacting magnon gas with conserved particle number emerges. Turbulence in this regime was already discussed in Ref.[\onlinecite{2016spinturbulence}]. Another example for statistically conserved $\hat{N}$ occur in the presence of strong anisotropy $\Delta $. In this case, $\hat{N}$ is approximately conserved due to suppression of longitudinal ($z$) spin fluctuations. This scenario will be discussed at the end. 

With these considerations in mind, the central quantity characterizing spin turbulence is the energy flux in $\bm k$ space. To define this quantity explicitly, we first express $\hat{H}$ in momentum space, $\hat{H} = \sum_{\bm k} J_{\bm k}\hat{\bm S}_{-\bm k}\cdot \hat{\bm S}_{\bm k}$, with $\hat{\bm S}_{\bm k} = \frac{1}{N}\sum_{i}\hat{\bm S}_i e^{-i{\bm k}\cdot{\bm r}_i}$ and $J_{\bm k} = J\sum_{\bm a} (1-e^{i{\bm k}\cdot{\bm a}})$ ($\bm a$ are unit-cell vectors and we already assumed the isotropic case $\Delta=1$ for simplicity). The 
energy flux in a momentum shell of radius $p$ can then be computed as 
\be
  \displaystyle  \Pi_E(p,t) = \sum_{|\bm k|<p} \sum_{\alpha=x,y,z}J_{\bm k} \frac{d{\cal C}_{\bm k}^\alpha}{dt}. 
\label{eq:Eflux}
\ee
In the definition (\ref{eq:Eflux}), we integrate the flux in the circular region $|\bm k|<p$ even though ${\cal C}_{\bm k}^\alpha$ is not radially symmetric in a lattice model such as ours. Because we focus on long wavelengths, the details of the area of integration do not affect our results. In the presence of emergent conserved quantities, we can also define corresponding fluxes in the same fashion, {\it e.g.}, $\Pi_N(p,t) = \sum_{|\bm k|<p} \frac{d({\cal C}_{\bm k}^x+{\cal C}_{\bm k}^y)}{dt}$ for $\hat{N}$ discussed above. 

\noindent{\bf Scaling of fluctuations through wave turbulence.}---Wave turbulence\cite{bookzakharov,booknazarenko} provides a framework for computing the scaling of two-point correlation functions when the system exhibits a weak coupling limit. We note that, far from the fully-polarized ground state, it is {\it a priori} unclear that such limit exists as the semiclassical equations of motion, the Landau-Lifshitz (LL) equations in (\ref{eq:eom}), are intrinsically non-linear ({\it i.e.}, there is no explicit weak coupling constant). Here we provide physical arguments to justify that such weak-coupling picture emerges even when the state has no net magnetization. The main assumption is that the correlation length $\xi $ is large, $a \ll \xi \ll L$ but not long-ranged, as expected for a two-dimensional ferromagnet [see Fig.\ref{fig:schematics}(c)]. Within a magnetic texture of size $\xi$, the fast, short-wavelength fluctuations ($|{\bm k}|\gtrsim 1/\xi$), responsible for transporting energy towards the UV, are incoherent. The starting point are the saddle point equations of motion of the spin-coherent path integral\cite{auerbachbook}, 
\be
i\partial_t {S}_i^\pm = J\sum_{j\in{\cal N}_i}\left[{S}_i^\pm {S}_j^z -{S}_j^\pm {S}_i^z\right],
\label{eq:eom}
\ee
with $S_i^\pm=S_i^x\pm iS_i^y$. Focusing on a small region of size $\xi$ in which magnetization is pointing in the $z$ direction, we replace $S_i^z \approx S - \frac{1}{2S}S_i^+S_i^-$ in Eq.(\ref{eq:eom}). As a result, the equations of motion for the short-wavelength spin fluctuations in ${\bm k}$-space are: 
\be
i\partial_t S_{\bm k}^+ = \omega_{\bm k} S_{\bm k}^+ +\sum_{{\bm k}_1,{\bm k}_2}V_{{\bm k},{\bm k}_1,{\bm k}_2} S_{\bm k-{\bm k}_1 - {\bm k}_2}^- S_{\bm k_1}^+S_{\bm k_2}^+,
\label{eq:classicaleom}
\ee
where $\omega_{\bm k} = JS\sum_{\bm a} (1-e^{i{\bm k}\cdot{\bm a}})$ and $V_{{\bm k},{\bm k}_1,{\bm k}_2} = \frac{J}{2S}\sum_{\bm a} \left[ e^{i{\bm k}_1\cdot{\bm a}} -  e^{i({\bm k}-{\bm k}_1)\cdot{\bm a}}\right]$. Equation\,(\ref{eq:classicaleom}) describes weakly coupled waves given that nonlinearities are smaller than linear terms by a factor $\langle|{\bm S}_i^+|^2\rangle/S^2 \ll 1$. 

The next step is to assume that transverse magnetization fluctuations are incoherent, {\it i.e.}, $\langle S_{\bm k}^\pm\rangle_S =0$ and $\langle S_{-\bm p}^-S_{\bm k}^+\rangle_S = \delta_{\bm p \bm k}n_{\bm k}$, with $S_{\bm k}^\pm$ the Fourier transform of $S_i^\pm$ and $n_{\bm k}$ the distribution function (here $\langle\ldots\rangle_S$ denotes sampling of $S_i^\alpha$ using a quantum distribution, see below). The standard procedure in wave turbulence consists of: (i) deriving a kinetic equation from (\ref{eq:classicaleom}) describing the time evolution of $\langle S_{-\bm k}^+S_{-\bm k}^-\rangle_S = n_{\bm k}$, (ii) proposing a solution of the form $n_{\bm k} \propto 1/|\bm k|^\nu$, and (iii) finding $\nu$ that gives rise to a steady-state solution. The exponent $\nu$ can only depend on the power $\alpha$ of the dispersion $\omega_{\bm k}\propto |\bm k|^\alpha$, the power $\beta$ of the interaction, $V({\lambda}{\bm k}_1,{\lambda}{\bm k}_2,{\lambda}{\bm k}_3,{\lambda}{\bm k}_4) = \lambda^\beta V({\bm k}_1,{\bm k}_2,{\bm k}_3,{\bm k}_4)$, and $d$. Inspection of the long wavelength limit of $\omega_{\bm k}$ and $V$ in Eq.(\ref{eq:classicaleom}) results in $\alpha=2$ and $\beta = 2$\cite{2019saraswat}. As shown in Refs.[\onlinecite{bookzakharov}] and [\onlinecite{booknazarenko}], there are two non-thermal solutions with scaling exponents
\be
\begin{array}{l}
  \nu_E = d+\frac{2\beta}{3} = \frac{10}{3},  \quad\quad
  \nu_{N} = d+\frac{2\beta}{3}-\frac{\alpha}{3} = \frac{8}{3},
  \label{eq:exponents}
  \end{array}
\ee
associated to a direct energy cascade ({\it i.e.}, flowing towards large momenta) and an inverse quasi particle cascade ({\it i.e.}, flowing towards small momenta), respectively. The forward energy cascade (reflecting the tranfer of energy from spin texture to quasiparticles) is a legitimate solution as it is consistent with our assumption of incoherent transport for scales $|\bm k| \gtrsim 1/\xi $. However, the inverse incoherent transport of quasiparticles is incompatible with the existence of magnetization textures at long wavelengths, which is confirmed in our numerics below [{\it i.e.}, relevant coherences in the IR are not captured by the incoherent fluctuation   approximation used in the derivation of Eq.(\ref{eq:exponents})]. When SU(2) symmetry is broken, suppresion of spin textures gives rise to a transverse magnetization cascade.

\begin{figure}
\includegraphics[scale = 1.0]{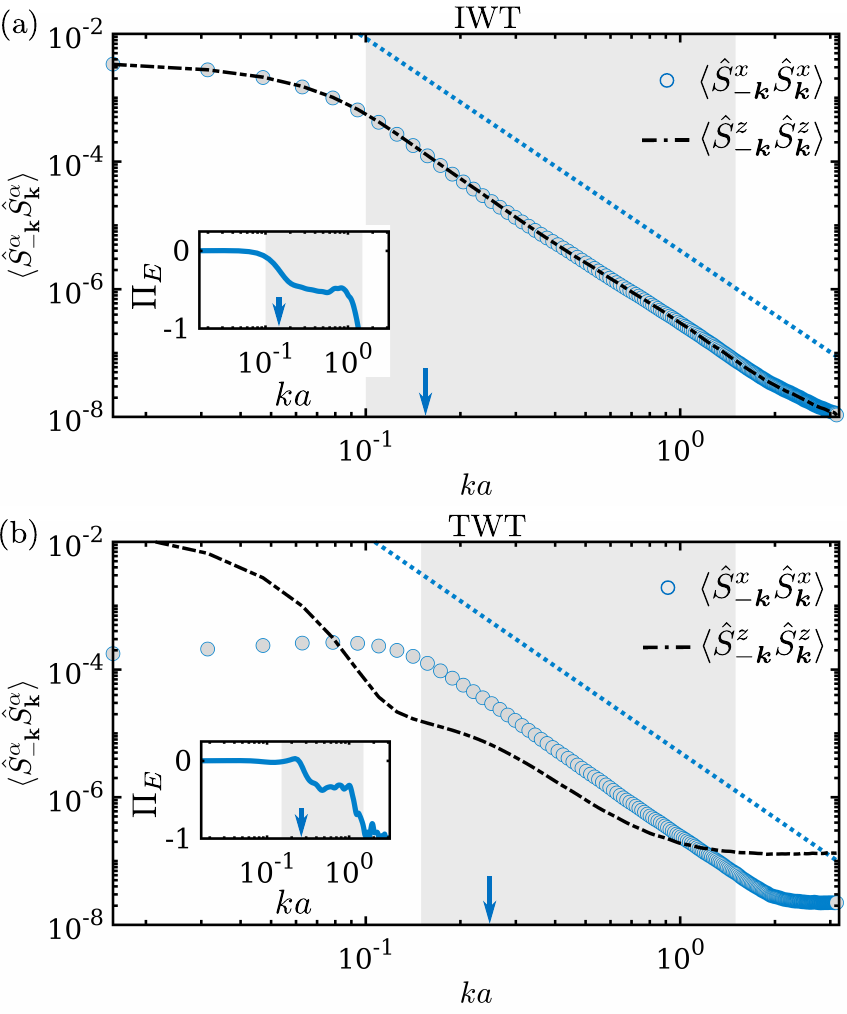}
\caption{Equal time spin-spin correlation function $\langle \hat{S}_{-\bm k}^\alpha(t)\hat{S}_{\bm k}^\alpha(t)\rangle$ after a quench computed via the Truncated Wigner Approximation for $t/\tau_* = 25$. Shown are simulation with (a) zero magnetization and (b) finite magnetization. Indicated with dotted lines is the energy cascade   power law $\sim 1/k^{\nu_E}$ [Eq.(\ref{eq:exponents})] and the shaded areas indicate the inertial range $1/\xi\lesssim |{\bm k}|\lesssim 1/a$. The energy flux $\Pi_E$ (see insets) exhibit a plateau in this range. Shown with arrows in the x-axis is the   wavevector ${\bm q}$ of the initial state.  Parameters used: (a) $q_xa =0.15$, $q_y = 0$, and $\theta = \pi/2$; (b) $q_xa =0.25$, $q_y=0$, and $\theta=\pi/6$. Parameters used: $L = 400$, and $S = 10$.}
  \label{fig:wt}
\end{figure}

\noindent{\bf Numerical simulations.}---To capture the short time scales in which the turbulent cascade develops, we use the Truncated Wigner Approximation\cite{spintwa1,spintwa2,spintwa3}. In this approximation, the semiclassical equations of motion (\ref{eq:eom}) are supplemented with quantum fluctuations drawn from a Wigner function. Defining $\langle \hat{\bm S}_i^\perp \rangle$ as the transverse magnetization with respect to the magnetization axis in Eq.(\ref{eq:initial}), we sample trajectories using Gaussian fluctuations of $\hat{\bm S}_i^\perp$ such that $\langle\hat{\bm S}_i^\perp\rangle = 0$ and $\langle\hat{\bm S}_i^\perp\cdot\hat{\bm S}_i^\perp\rangle = S$. Once Gaussian fluctuations are included, connected correlations become finite [{\it i.e.}, $\langle \hat{S}_i\hat{S}_j\rangle \neq \langle \hat{S}_i\rangle \langle\hat{S}_j\rangle$].

Figure\,\ref{fig:wt} shows the spin-spin correlation function after the spin spiral order has been destroyed, which occurs on a timescale $t/\tau_*\approx 5$\cite{2020spinspiral}. In panel (a), we consider an initial state with zero net magnetization ($\theta=\pi/2$) and $q_xa = 0.15$. In this case, although the initial state is anisotropic, the dynamic instability restores the rotational symmetry and all components of magnetization exhibit the same scaling behavior. In our simulations, we observe the development of a single power law at wavevectors $|{\bm k}|\gtrsim |{\bm q}|$, see shaded area. The observed power law is consistent with an energy cascade ($\nu_E\approx 3.33$), which can be confirmed by showing that the energy flux $\Pi_E$ exhibits a plateau in the inertial range (see inset). For an initial state with finite magnetization [Fig.\ref{fig:wt}(b)], there is a remaining anisotropy between the transverse and longitudinal fluctuations after the spiral order has been destroyed. In this case, only the tranverse magnetization exhibit scaling with the same characteristics as the ones described in panel (a). 

We do not observe the scaling $1/|{\bm k}|^{\nu_N}$ in any of our simulations for the isotropic Heisenberg model. Indeed, we observe that spin textures dominate the infrared at all times, as shown in Fig.\ref{fig:schematics}(c), therefore impeding the inverse (incoherent) quasiparticle cascade. 

We emphasize two important points. First, the turbulent scaling is insensitive to the details of the initial condition and the same qualitative behavior occurs, for instance, with an incoherent initial state. In this case, turbulence appears in a much shorter timescale $t/\tau_{*}\approx 1$. Second, the lack of wave turbulence for large energy densities does not imply absence of turbulence. Indeed, for a high energy spin spiral it can be shown that a (nonuniversal) energy cascade can still form. We provide numerical evidence of these two points in the Supplement. 

\begin{figure}
\includegraphics[scale = 1.0]{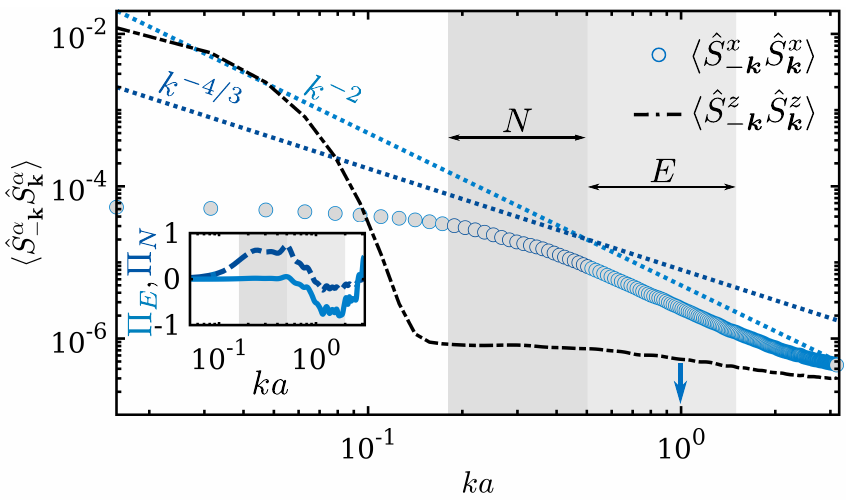}
\caption{Spin-spin correlation function shown for the XXZ model. A dual cascade, indicated with two shades of gray, is observed. The scaling at large $\bm k$ is consistent with an energy cascade ($\nu_E = 2$), whereas the scaling at smaller $\bm k$ is consistent with a particle cascade ($\nu_N = 4/3$). The inset shows the energy and particle fluxes in ${\bm k}$-space. Parameters used: incoherent initial conditions with $q_xa = 1$, $q_y=0$, $\theta = \pi/4$, $t/\tau_{*} = 15$, $\Delta=0.5$.}
  \label{fig:xxz}
\end{figure}

\noindent{\bf Quantum fluctuations.}---We analytically estimate the width of the inertial range (the equivalent of the Reynold's number in hydrodynamics) by finding the wavevector $q_*$ above which we expect quantum dissipation to be important. We use a simple scaling argument that employs the exact spin operator identity $\frac{1}{N}\sum_{\bm k,\alpha}\hat{S}_{-\bm k}^\alpha\hat{S}_{-\bm k}^\alpha = S(S+1)$. At the onset of turbulence, we anticipate that all mean-field values $\langle\hat{S}_{\bm k}^\alpha\rangle$ relax to zero for $\bm k \neq 0$ due to dephasing, and $\langle \hat{S}_{\bm k = 0}^{\alpha}\rangle = \sqrt{N} S \cos\theta \delta_{\alpha,z}$ due to spin conservation. The central idea in our argument is to express $\sum_{\alpha}\langle \hat{S}_{-\bm k}^\alpha\hat{S}_{\bm k}^\alpha\rangle = \chi_{\bm k}^{\rm c}+\chi_{\bm k}^{\rm q}$, where $\chi_{\bm k}^{\rm c}$ accounts for the `classical' weight, $\frac{1}{N}\sum_{\bm k}\chi_{\bm k}^{\rm c} = S^2\sin^2\theta$, and $\chi_{\bm k}^{\rm q}$ accounts for the `quantum' weight $\frac{1}{N}\sum_{\bm k}\chi_{\bm k}^{\rm q} = S$. We first assume that $\chi_{\bm k}^{\rm c}$ takes a simple form consistent with a turbulent cascade: $\chi_{\bm k}^{\rm c} = A$ for $|{\bm k}|<q$, and $\chi_{\bm k}^{\rm c} = A(q/|{\bm k}|)^{\nu_E}$ for $|{\bm k}|>q$, with $A = \frac{4\pi(\nu_E-2)S^2\sin^2\theta}{\nu_E(qa)^2}$. In addition, we assume that $\chi_{\bm k}^{\rm q} $ is uniform in $\bm k$ space, {\it i.e.}, $\chi_{\bm k}^{\rm q}=S$. As a result, the wavevectors at which quantum fluctuations become important can be obtained from the condition $\chi_{\bm k}^{\rm c} = \chi_{\bm k}^{\rm q}$. This condition yields an inertial range of size $\frac{q_*}{q}= \left[\frac{4\pi(\nu_E-2)}{\nu_E}{\cal Q}\right]^{1/\nu_E}$ which is controlled by the parameter ${\cal Q}$ in Eq.(\ref{eq:reynolds}) (note that there is also physical lattice cutoff $q_* \le 1/a$). In particular, for $\theta = \pi/2$, we find numerically that ${\cal Q} \gtrsim 60$ is needed for the observation of a clear power-law energy cascade (see Supplement). 

\noindent{\bf The Heisenberg XXZ model.}---The computation of scaling exponents in the anisotropic Heisenberg model proceeds in the same way as in the isotropic case. The main difference is that the interaction $V$ becomes wavevector independent, $\beta = 0$\cite{2019saraswat}. In this case, the predicted wave turbulence exponents are  $\nu_E = 2$ and $\nu_N = 4/3$. As shown in Fig.\ref{fig:xxz}, the simulations exhibit a qualitatively distinct behavior with respect to the isotropic case. We first note that longitudinal fluctuations $\langle \hat{S}_{-\bm k}^z\hat{S}_{\bm k}^z\rangle$ (dashed-dotted lines) are strongly suppressed within the inertial range, which justifies the emergence of a conserved $\hat{N}$ ({\it i.e.}, turbulence is effectively occurring in the transverse magnetization sector). Second, we observe that the power-law exponent associated to the energy cascade ($\nu_E$=2) appears only in a small sector of the inertial range. In fact, a second cascade with an exponent consistent with a quasiparticle cascade ($\nu_N = 4/3$) is found to dominate the inertial range. We confirm the dual nature of the cascade by plotting the energy and quasiparticle fluxes in $\bm k$-space, see inset (note that the energy flux vanishes where the particle flux is finite and constant, and viceversa).

\noindent{\bf Connection to experiments.}---The far-from-equilibrium dynamics of spin systems have recently been probed in a variety of quench experiments in cold atoms\cite{2014spiralscience,2014spiralexp,2020spiralketterle}, making our predictions within experimental reach. Indeed, experiments can already access relatively long timescales $t\sim 40\frac{1}{J}$, sufficient to capture the onset of the turbulent cascade. Our results are also relevant in the context of driven ferromagnetic insulators, such as YIG\cite{2017chunhui}: a turbulent cascade can be generated by driving the system at the ferromagnetic resonance, and the turbulent spectrum at large energies can be measured through noise magnetometry\cite{2018nv-wire,2018nv-ferro} (however, more detailed studies about the effects of dipolar interactions and dissipation to the lattice are required in this case). 

\noindent{\bf Concluding remarks.}---The turbulent scaling in the isotropic and XXZ Heisenberg models is insensitive to microscopic details and emerges for a broad range of initial conditions, either coherent or incoherent, far from the ferromagnetic ground state. This suggests that signatures of spin turbulence may be predominant and readily accessible in various experimental platforms. Open problems that remain to be addressed are studying the emergence of spatial/temporal scaling after the quench and understanding the interplay between thermal and quantum fluctuations at long times\cite{2006becdissipation,2007qturbulencedecayexp,2017turbulencedecay}. In addition, the possibility to engineer the spin-spin interaction in cold atom platforms motivates the study of turbulence in different settings, such as dipolar gases with long range interactions. 

\vspace{5mm}

{\bf Acknowledgements.}---JFRN thanks Eugene Demler, Gregory Falkovich, Sean Hartnoll, Vedika Khemani, Jamir Marino, Xiaoliang Qi, and Monika Schleier-Smith for insightful comments and discussions, and specially Paolo Glorioso, Wen Wei Ho, and Asier Pi{\~ n}eiro-Orioli for a critical reading of the manuscript. JFRN is supported by the Gordon and Betty Moore Foundation's EPiQS Initiative through Grant GBMF4302 and GBMF8686. JFRN also acknowledges the 2019 KITP program {\it Spin and Heat Transport in Quantum and Topological Materials} and the National Science Foundation under Grant No. NSF PHY-1748958.

%

\clearpage

\renewcommand{\thefigure}{S\arabic{figure}}
\renewcommand{\theequation}{S\arabic{equation}}
\renewcommand{\thesection}{S\arabic{section}}
\setcounter{page}{1}
\setcounter{equation}{0}
\setcounter{figure}{0}
\setcounter{section}{0}

\begin{widetext}

\begin{center}
{\large\bf Supplement for `Turbulent relaxation after a quench in the Heisenberg model'}

\vspace{4mm}

Joaquin F. Rodriguez-Nieva

{\small\it Department of Physics, Stanford University, Stanford, CA 94305, USA}

\end{center}


\vspace{6mm}

The Supplement provides additional numerical results of turbulence in different regimes. In Sec.S1, we discuss turbulence obtained from evolving an incoherent initial state, different from the coherent spin spiral discussed in the main text. In Sec.S2, we discuss the non-universal energy cascade in the large energy density regime. In Sec.S3, we study the sensitivity of the scaling exponents as a function of the spin number $S$. 

\section{S1. Incoherent initial conditions}

The spin spiral state is a typical state in cold atom experiments. In solid state systems, however, it is also common to incoherently drive the system with an oscillating magnetic field. As we show next, the scaling exponents are insensitive to the details of the initial state.

We consider an incoherent initial condition parametrized as $\langle \hat{S}_i^+\rangle = \sum_{\bm k}f_{\bm k}e^{i{\bm k}\cdot{\bm r}_i}$. To do a one-to-one comparison with the numerics of the main text, we consider that the distribution $|f_{\bm k}|^2$ is Gaussian and peaked at the wavevector $\bm q$, and the value of $\sum_{\bm k}|f_{\bm k}|^2$ defines the total magnetization of the initial incoherent state. The numerical results are shown in Fig.\ref{fig:incoherent}(a) for a state with finite magnetization in the $z$ direction. Compared to Fig.\ref{fig:wt}(b) of the main text, the qualitative agreement between both results is notable: the power law cascade is formed at the wavevector ${\bm q}$ (indicated with an arrow), and the same power law scaling (associated to an energy cascade) is observed. The energy flux also exhibits a plateau in the inertial range. 

\begin{figure*}[b]
\includegraphics[scale = 1.0]{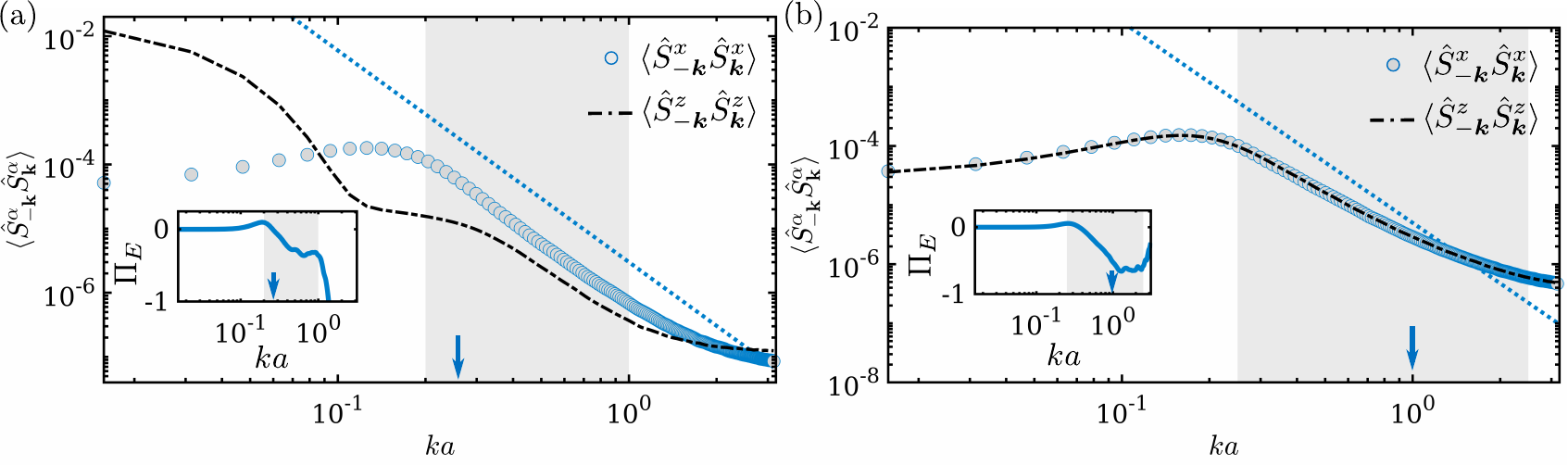}
\caption{(a) Equal time spin-spin correlation function obtained from quenching an initial incoherent state using the Truncated Wigner Approximation, see discussion in Supplementary text. The initial condition has wavevector $q_xa = 0.25$, $q_y=0$, and finite magnetization in the $z$ axis ($\theta = \pi/4$). (b) Equal time spin-spin correlation function obtained from quenching an initial spin spiral state with large energy density. Parameters used: $q_xa = 1$, $q_y=0$, $\theta = \pi/2$. In both panels, the shaded area indicates the inertial range. The insets show the energy flux in $\bm k$-space.} 
  \label{fig:incoherent}
\end{figure*}

\section{S2. Beyond wave turbulence}

The wave turbulence exponents are not valid when the energy density is large. In this case, the picture of wave turbulence developing within large magnetization texture breaks down. Figure\,\ref{fig:incoherent}(b) shows a simulation in which the initial state has a large energy density, ${q}_xa=1$ and $\theta = \pi/2$. Clearly, in this case there is no power law scaling of the distribution function. Indeed, because the initial wavevector is imprinted at intermediate timescales, the long wavelength expansion to describe wave turbulence breaks down. Nonetheless, it is still possible to define an energy cascade at large momenta (see inset).

\section{S3. Limit of small spin number $S$}

In the main text, we estimated the effects of quantum statistics on the short wavelength physics. We argued that, if ${\cal Q}\gg 1$ [see Eq.(\ref{eq:reynolds})], then we expect an energy cascade consistent with the exponents derived from the Landau-Lifshitz equation. Here we discuss the effects of quantum statistics on the power law scaling in the limit of small $S$, which is the case of relevance in most experiments. 

Figure \ref{fig:Sdependence} reproduces the conditions studied in Fig.\ref{fig:wt}(a) of the main text for different values of $S$ and $q_x$ (in all cases, we used $\theta = \pi/2$). For $q_xa = 0.13$ [panel (a)], we observe a clear energy cascade ($\nu_E = 10/3$) for arbitrary $S$. For $q_xa = 0.25$ [panel (b)], we observe sizable deviations from the power-law turbulent cascade in the UV for $S<4$. Consequently, the transition from turbulence to quantum dissipative occurs on ${\cal Q}_* \approx 60$. 

\begin{figure}
\includegraphics[scale = 1.0]{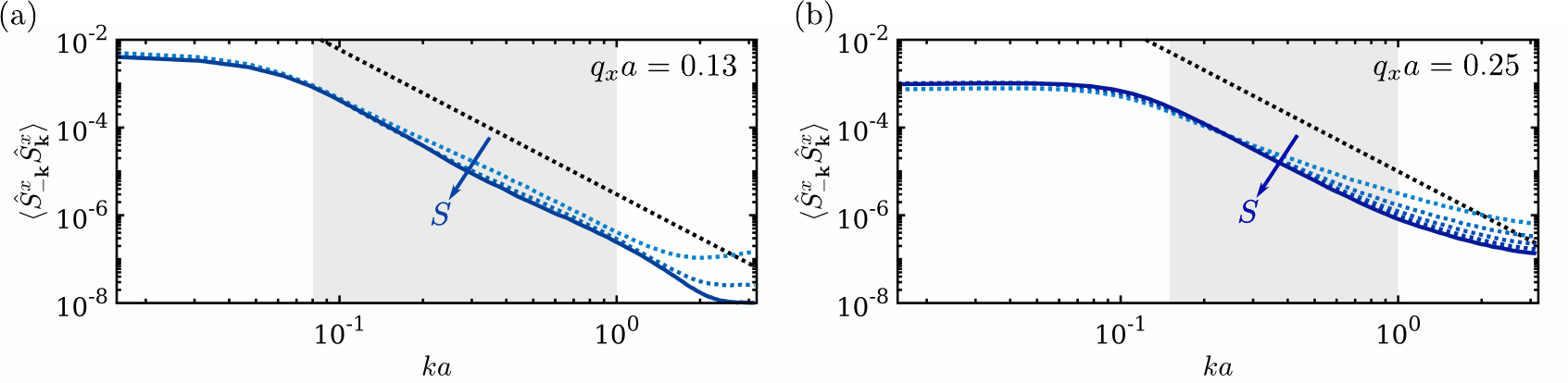}
\caption{Spin-spin correlation function for different values of the spiral wavevector $q$ and spin number $S$. The initial state is the same as that in Fig.\ref{fig:wt}(a) of the main text. Indicated with decreasing shades of blue is the results for decreasing $S$, where we used $S = 1,3,5$ in (a) and $S = 2,4,6,8,10$ in (b).}
  \label{fig:Sdependence}
\end{figure}

\end{widetext}

\end{document}